\documentclass[preprint,superscriptaddress,floatfix]{revtex4}

\usepackage{amsmath}
\usepackage{bm}
\usepackage{amsfonts}
\usepackage[pdftex]{color,graphicx}

\newcommand{\indTab}
{
  \begin{table}[b]
    \caption{Key parameters of parallelizable loops in DFT calculation and their scales, considering usual large-scale calculation.} \label{Tab1}
    \begin{ruledtabular}
      \begin{tabular}{cc}
        Parameter  & Loop scale  \\
        \hline
        k-points$^a$  & O(1) $\sim$ O(100) \\
        Atoms  & O(100) $\sim$ O(1000) \\
        Basis functions & O(1000) $\sim$ O(10000) \\
        Auxiliary basis functions$^a$ & O(100000) $\sim$ O(1000000)  \\
      \end{tabular}
    \end{ruledtabular}

    $^a$ If their arrays are one-dimensionalized.
  \end{table}
}

\newcommand{\tetTab}
{
  \begin{table}[b]
    \caption{Computation time (millisecond) for normalization of n vectors of n complex number elements.  Each value is the average of 10 time calculation.} \label{Tab2}
    \begin{ruledtabular}
      \begin{tabular}{ccccccccc}
        N & Serial & \multicolumn{2}{c}{OpenMP (4 threads)} & \multicolumn{2}{c}{OpenMP (8 threads)} & GPU$^a$ & { } & Transfer$^b$  \\
        { } & { } & min & max & min & max & min & max & { }  \\
        \hline
        50 & 0.01 & 0.03 & 0.05 & 0.05 & 0.07 & 0.08 & 0.19 & 0.08  \\
        100 & 0.05 & 0.07 & 0.13 & 0.09 & 0.16 & 0.17 & 0.40 & 0.23  \\
        500 & 1.87 & 0.57 & 2.33 & 0.73 & 1.89 & 0.96 & 2.47 & 3.44  \\
        1000 & 9.51 & 2.50 & 13.11 & 2.17 & 6.01 & 1.80 & 5.58 & 13.09 \\
        2000 & 49.89 & 10.55 & 60.55 & 8.67 & 26.17 & 4.38 & 19.52 & 49.62 \\
        3000 & 138.94 & 22.80 & 153.20 & 19.81 & 63.82 & 7.38 & 47.34 & 110.46 \\
      \end{tabular}
    \end{ruledtabular}

    $^a$ Excluding the transfer time between a CPU and a GPU \\
    $^b$ Transfer time between a host CPU and a GPU device
  \end{table}
}

\newcommand{\rouTab}
{
  \begin{table}[b]
    \caption{Time sharing of typical 1 CPU DFT calculation on the magnetite system by OpenMX.} \label{Tab3}
    \begin{ruledtabular}
      \begin{tabular}{ccc}
        Routine & Time (second) & Share(\%)  \\
        \hline
        Set OLP Kin & 287.037 & 0.130 \\
        Set Nonlocal & 872.376 & 0.394 \\
        Set Hamiltonian & 7091.817 & 3.201 \\
        Poisson & 3.429 & 0.002 \\
        Diagonalization & 208811.550 & 94.236 \\
        Mixing DM & 62.969 & 0.028 \\
        Force & 1433.113 & 0.647 \\
        Total Energy & 299.321 & 0.135 \\
        Set Aden Grid & 2.011 & 0.001 \\
        Set Orbitals Grid & 6.170 & 0.003 \\
        Set Density Grid & 2705.252 & 1.221 \\
        Others & 7.894 & 0.004 \\
      \end{tabular}
    \end{ruledtabular}
  \end{table}
}

\newcommand{\parTab}
{
  \begin{table}[b]
    \caption{Percentage of the routines to consume serial calculation CPU time in Band\_DFT\_Col.} \label{Tab4}
    \begin{ruledtabular}
      \begin{tabular}{cc}
        Routine & Share(\%)  \\
        \hline
        Imposing k dep. (Part 1) & 0.224 \\
        Diagonalizing S (Part 2) & 39.531 \\
        Making S$^\dagger$HS (Part 3) & 15.023 \\
        Daigonalizing S$^\dagger$HS (Part 4) & 39.390 \\
        Chem. Pot. \& Band E. (Part 5) & 0.001 \\
        Density Matrix (Part 6) & 5.788 \\
        Sum & 99.956 \\
      \end{tabular}
    \end{ruledtabular}
  \end{table}
}

\newcommand{\oneTab}
{
  \begin{table}[b]
    \caption{Speedup of Band\_DFT\_Col by each parallelization scheme using 1 quad-core CPU and 1 GPU, compared to serial calculation, and CPU memory usage} \label{Tab5}
    \begin{ruledtabular}
      \begin{tabular}{ccc}
        Parallelization scheme & Speedup & Host memory  \\
        \hline
        MPI & 3.73 & 6382 MB \\
        OpenMP & 1.52 & 3126 MB \\
        CUDA & 3.78 & 3126 MB \\
        CUDA \& OpenMP & 3.79 & 3126 MB \\
        MPI \& CUDA & 5.60 & 6382 MB \\
      \end{tabular}
    \end{ruledtabular}
  \end{table}
}

\newcommand{\EHHTab}
{
  \begin{table}[b]
    \caption{Percentage of the partial routines to consume serial calculation CPU time in Eigen\_HH and speedups in each routine by 1 GPU + OpenMP (4 threads) parallelization.} \label{Tab6}
    \begin{ruledtabular}
      \begin{tabular}{ccc}
        Routine & Percentage (\%) & Speedup  \\
        \hline
        Householder tridiagonalization & 43.17 & 5.23 \\
        LAPACK routine & 14.75 & 1.00 \\
        Rearrangement of eigenvectors & 42.02 & 4.23 \\
        Normalization \& transposition & 0.07 & 1.84 \\
      \end{tabular}
    \end{ruledtabular}
  \end{table}
}

\newcommand{\twoTab}
{
  \begin{table}[b]
    \caption{Speedup of Band\_DFT\_Col by each parallelization scheme using 2 quad-core CPUs and 2 GPUs, compared to serial calculation, and CPU memory usage as well as speed increase rate by processor doubling, i.e., by adding 1 CPU and 1 GPU. OpenMP used 4 threads in each MPI process.} \label{Tab7}
    \begin{ruledtabular}
      \begin{tabular}{cccc}
        Parallelization scheme & Speedup & Host memory & Increase rate \\
        \hline
        MPI & 6.33 & 10683 MB & 70 \% \\
        MPI \& OpenMP & 2.89 & 4211 MB & 90 \% \\
        3-way hybrid & 7.55 & 4211 MB & 99 \% \\
        MPI \& CUDA & 10.94 & 10683 MB & 95 \% \\
      \end{tabular}
    \end{ruledtabular}
  \end{table}
}

\newcommand{\hybFig}
{
\begin{figure}[t]
   \centering
   \includegraphics[width=\columnwidth]{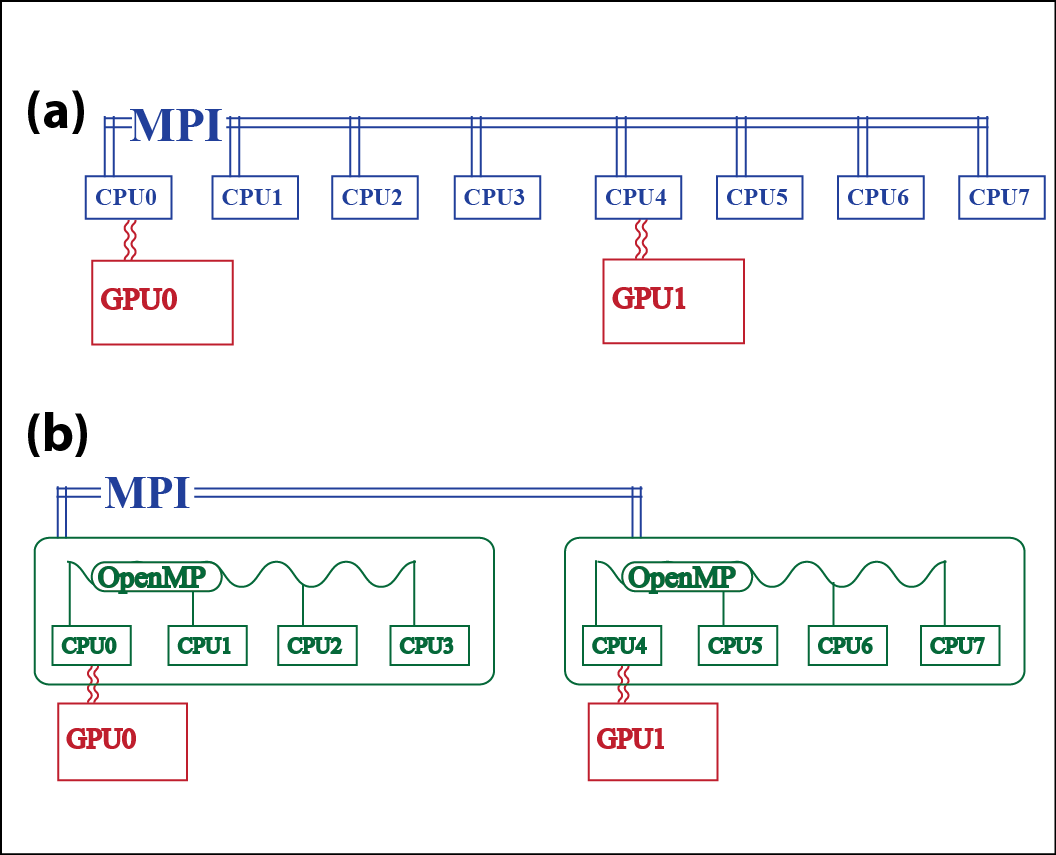}
   \caption{Schematic pictures of hybrid parallelization candidates of (a) MPI+OpenMP+CUDA and (b) MPI+CUDA.}
   \label{Fig1}
\end{figure}
}

\newcommand{\geoFig}
{
\begin{figure}[t]
   \centering
   \includegraphics[width=\columnwidth]{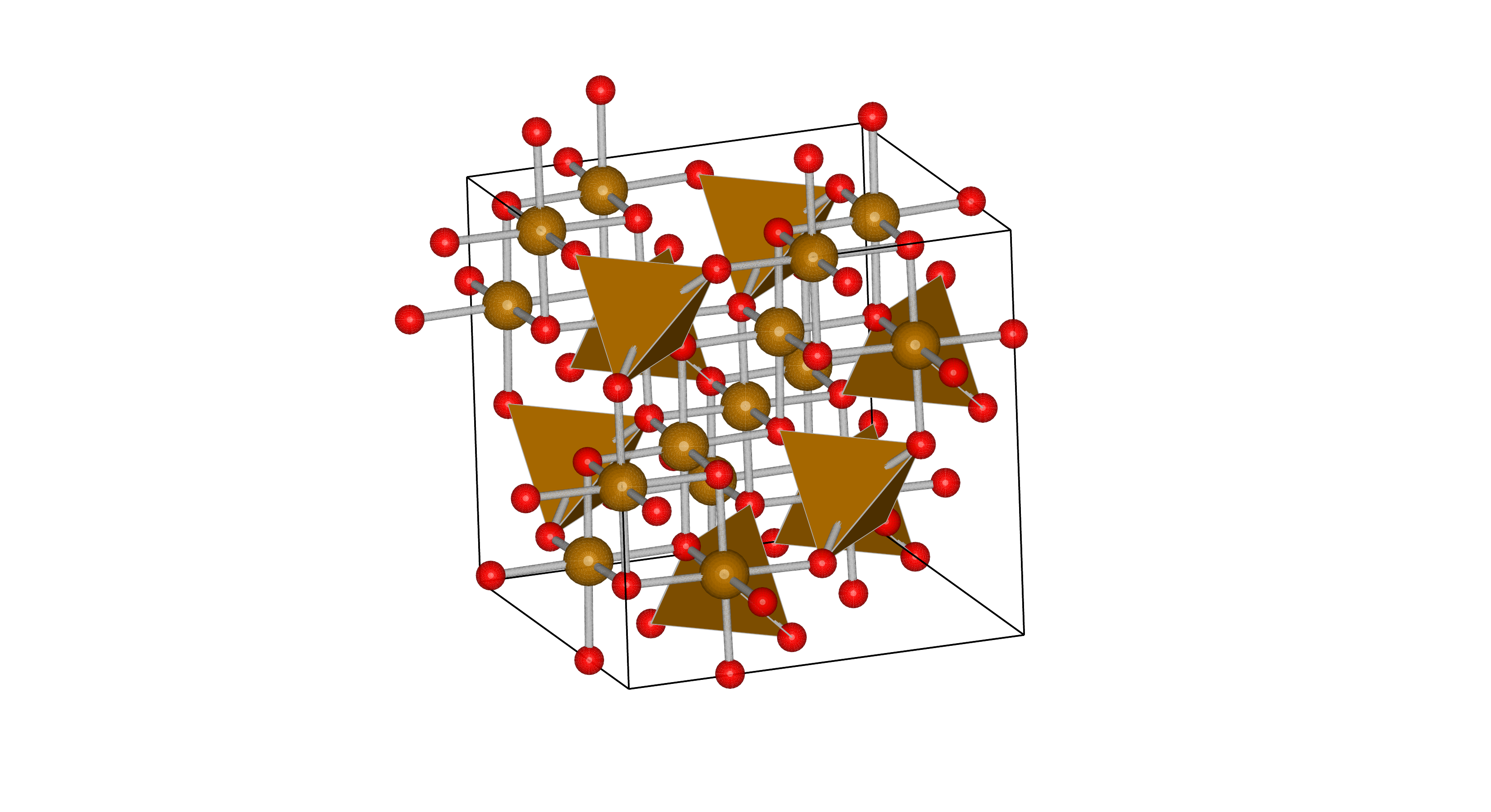}
   \caption{3D view~\cite{31} of cubic crystal stucture of our magnetite system.}
   \label{Fig2}
\end{figure}
}

\newcommand{\parFig}
{
\begin{figure}[t]
   \centering
   \includegraphics[width=\columnwidth]{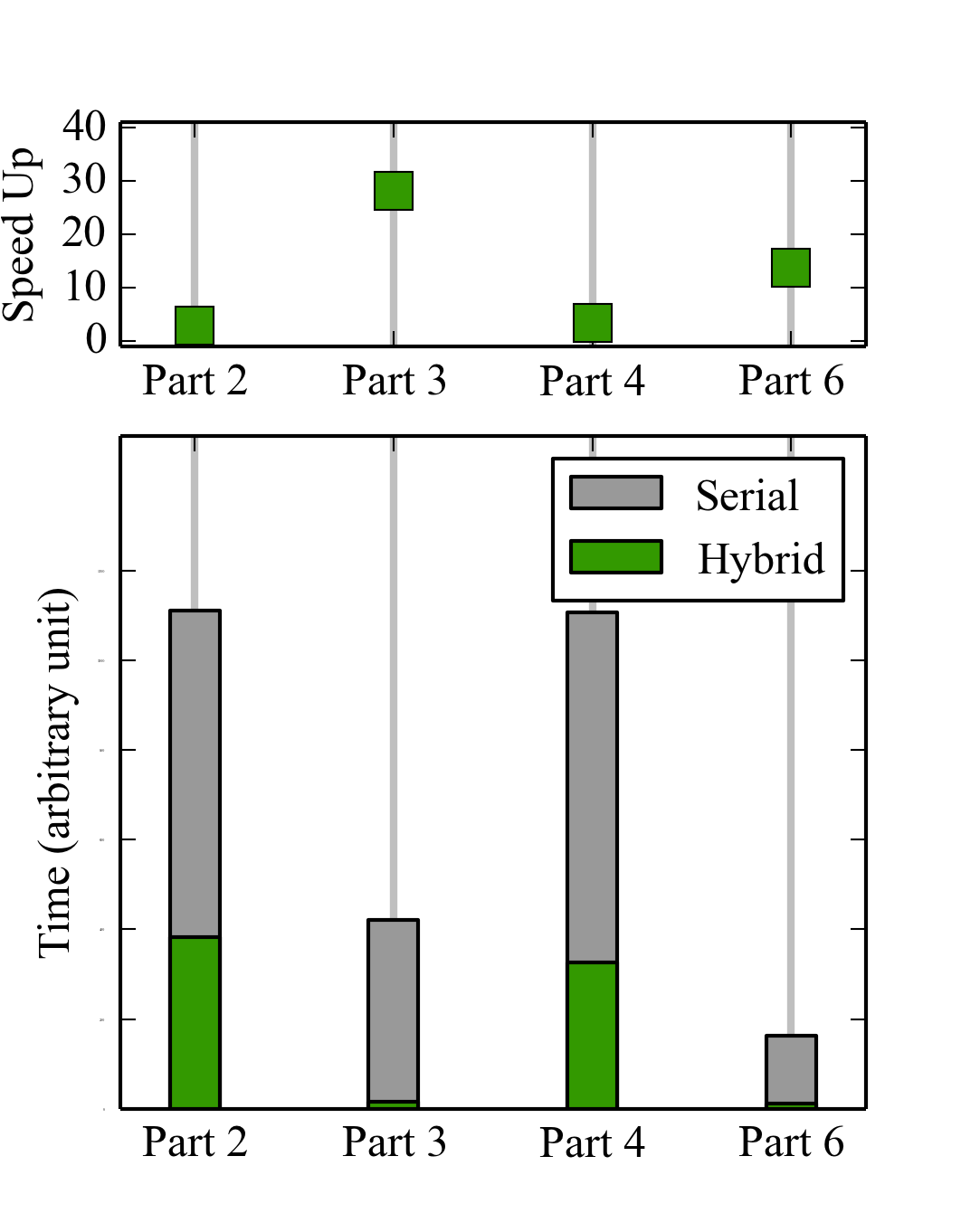}
   \caption{Time and speedup for various principal routines, by 1 quad-core CPU and 1 GPU within OpenMP+CUDA parallelization (denoted Hybrid).}
   \label{Fig3}
\end{figure}
}

\begin{document}

\title{Effects of Easy Hybrid Parallelization with CUDA for Numerical-Atomic-Orbital Density Functional Theory Calculation}

\author{Jae-Hyeon Parq}
\email[Electronic mail: ]{parkq2@snu.ac.kr}
\affiliation{QoLT IIDC and School of Earth and Environmental Sciences, Seoul National University, Seoul 151-747, Korea}

\author{Erik Sevre}
\affiliation{QoLT IIDC and School of Earth and Environmental Sciences, Seoul National University, Seoul 151-747, Korea}

\author{Sang-Mook Lee}
\affiliation{QoLT IIDC and School of Earth and Environmental Sciences, Seoul National University, Seoul 151-747, Korea}

\begin{abstract}

We modified a MPI-friendly density functional theory (DFT) source code within hybrid parallelization including CUDA.  Our objective is to find out how simple conversions within the hybrid parallelization with mid-range GPUs affect DFT code not originally suitable to CUDA.  We settled several rules of hybrid parallelization for numerical-atomic-orbital (NAO) DFT codes.  The test was performed on a magnetite material system with OpenMX code by utilizing a hardware system containing 2 Xeon E5606 CPUs and 2 Quadro 4000 GPUs.  3-way hybrid routines obtained a speedup of 7.55 while 2-way hybrid speedup by 10.94.  GPUs with CUDA complement the efficiency of OpenMP and compensate CPUs' excessive competition within MPI.
 
\end{abstract}

\maketitle

\section{Introduction}
 The density functional theory (DFT) is a quantum calculation tool widely used in material sciences.  Compared to other quantum chemical methods, DFT prides itself on its low computational cost and moderate accuracy~\cite{1}, which make DFT useful in recent theoretical research of large model systems consisting of many atoms.  There are various DFT code packages classified by wavefunction basis sets; OpenMX~\cite{2} and SIESTA~\cite{3} are famous as code packages using numerical-atomic-orbital (NAO) basis sets.  The keys to how DFT codes have succeeded in the last decades are adequate approximations and suitable parallelization with message-passing interface (MPI)~\cite{4}.

Parallelization methods for newly developed computer architectures are becoming more important since research simulations need increasingly large model sizes.  As high-end computer architectures have adopted cluster structures with memory distributed in nodes, MPI has been dominantly used due to its convenience to users.  Development of multi-core architectures, however, has increased utilization of OpenMP~\cite{5}, parallelism within shared memory systems.  Recent high performance computing architectures have applied heterogeneous memory systems and inspired hybrid parallelism using both MPI and OpenMP~\cite{6}.

Graphical processor units (GPUs) are improving so as to overcome the speed limit of CPUs in the name of general purpose computing on GPUs (GPGPU) although cache-based CPUs are still very successful for speeding up the high-end supercomputers.  The spirit of GPGPU is that GPUs assist CPUs and reduce total calculation time by converting some parts of an original CPU code into those optimal on GPUs.  The early stage of GPGPU suffered from poor programmability and lack of double-precision compute capability.  Programming for GPUs became easy with the introduction of NVIDIA's Compute Unified Device Architecture (CUDA)~\cite{7}, and the hardware problem disappeared with the release of NVIDIA's state-of-the-art GPUs with double-precision instruction sets. 

CUDA implementation has successfully accelerated DFT codes on plane-wave basis set~\cite{8} and on Gaussian basis set~\cite{9}.  In the former~\cite{8}, it is sufficient to utilize libraries distributed by NVIDIA and to rectify some code because the original CPU version code has full of easily converted parts.  If it is not the case, the code should be greatly transformed to be fit for GPU.  In the latter~\cite{9}, the original code's algorithm was changed to increase the efficiency on GPUs. 

Each parallelism has its own advantages and disadvantages.  MPI processes cannot share memory to require data transfer among processes.  OpenMP and CUDA apply shared memories but must follow established forms in memory access to reduce conflicts among threads, which would occur less within MPI.  CUDA's available threads outnumber OpenMP's, but a GPU device needs to transfer data with its host (CPUs).  Neither CUDA nor OpenMP can combine computer nodes, but MPI can.  Hybridization between these parallelisms means various kinds of parallelization need to undertake the difficult task to be combined in parts into a determined form, and there have been several successes of the hybrid parallelization with CUDA~\cite{10,11,12,13,14,15,16}.

Most of DFT codes are known to be MPI-friendly, that is, MPI parallelism has a good efficiency in them although that does not mean DFT codes are inappropriate to OpenMP or CUDA parallelism.  For DFT calculation, OpenMP is meaningful in economy of memory.  CUDA has nothing to do with CPU memory, but more CPU cores with MPI may be the better choice for DFT codes if a few GPUs with CUDA cannot provide a considerable benefit.  However, hybridization with CUDA can be attractive if it gives good synergy with existing code.

From a programmers' point of view, CUDA implementation requires considerable effort to obtain the best efficiency if the original code's algorithm is not suitable to GPGPU.  In that case, changing the essence of the original algorithm can be a big problem, editing the most of code.  Instead of all this work, hybrid parallelization with CUDA can be a compromise.  The problem is whether the speedup from the hybrid parallelization is comparable to that from changing the main algorithm.  A test for the hybrid parallelization without changing the main algorithm is therefore needed. 

This paper discusses strategies for hybrid parallelization on NAO DFT, and tests them to analyze their benefit.  Parallelizable parts in DFT codes have certain specific patterns, varied by basis set.  The next section argues on the ways of combination of MPI, OpenMP and CUDA for NAO DFT calculation, by examining such patterns.  In the following section, we applied the hybrid strategies to OpenMX package~\cite{code} and judge whether the resulting speedup is meaningful.

\section{Hybrid parallelization strategies for NAO DFT}
We first consider schematic hybrid parallelization in terms of hardware.  To utilize all CPUs and GPUs, we naturally consider combining all three methods, OpenMP, MPI, and CUDA parallelization.  When the number of GPU devices is less than that of CPU cores, one can divide the GPU devices to evenly match with the CPU cores.  Each (OpenMP) master thread communicates with its assigned GPU device via CUDA and with other master threads via MPI, as described in Fig. 1 (a).  This is similar to those suggested by some papers~\cite{12,13,14,15}, but there are small differences in them such that GPU execution does not overlap CPU operation or all CPU cores do not participate in OpenMP part works.  In our scheme, if the number of CPU cores is four times to that of GPU devices, for example, OpenMP should bind every 4 CPU cores, connected to every single GPU within CUDA, and MPI connects each bunch of 4 CPU cores with the other CPU core bunches.  This 3-way hybrid parallelization consumes less memory than parallelization without OpenMP since memory cost depends on the number of MPI processes, but it is not guaranteed that the resulting speed will surpass MPI only parallelization.
Another way of the hybrid parallelization is excluding OpenMP to obtain maximum speed and can be found in some literature~\cite{10}.  This type of hybrid parallelization will be demanded if the speed of the above 3-way hybrid parallelization fails to excel MPI only parallelization or if the contribution of OpenMP in the 3-way hybrid parallelization is too small.  All CPU cores communicate with each other via MPI as shown in Fig. 1 (b).  Only specific CPU cores are connected to GPU devices, i.e., nonequivalent allocation.  Thus this parallelization scheme guarantees the speed higher than MPI only parallelization.  In this case, however, load balancing among MPI processes is not trivial.  The best way of distributing load to CPU cores is probably dynamic load balancing or determination by tests.

\hybFig

The next is investigation in terms of software, i.e., examination of DFT codes from the viewpoint of hybrid parallelization.  The arguments here for parallelizing DFT codes are focused on 3-way hybrid method.  In the case of MPI+CUDA 2-way hybrid method, the parts to be assigned to OpenMP in 3-way hybrid method may be changed into CPU serial (which can be parallelized by MPI) or CUDA codes.  The problem of choosing whether to implement serial or CUDA is not trivial, but it seems related to the problem of choosing whether to implement OpenMP or CUDA. 

Large-scale parallelization usually concerns loops in a serial code, i.e., data parallelization, but task parallelization, requiring fixed number of processes or threads, is usually inappropriate for general cases of DFT codes, although OpenMP can handle it.  Thus the priority is checking what loops are contained in a DFT code.  Simply speaking, DFT calculation has a procedure of five parts~\cite{17}, initial setting, calculating effective potential, diagonalization (solving eigenvalue equation), calculating densities and obtaining results.  The middle three parts consist of the main iterated loop and spend the most of total calculation time.  So we will focus on these three parts.  The part to determine effective potential contains loops to calculate elements of the Hamiltonian matrix and the overlap matrix, both of which consist of the eigenvalue equation concerned by the diagonalization part.  The charge density or spin density is calculated within loops for density grid.

Before examining such loops, we check the indices of parallelizable loops in a DFT code.  Since iterated loops are not parallelizable, most of the time-consuming parallelizable loops in a DFT code appear to be related to the Hamiltonian matrix or the overlap matrix, both of which contain indices of spin, k-point and basis function.  In the case of NAO DFT, auxiliary basis functions~\cite{18} are added to calculate charge density or spin density as well as the matrix elements.  Finally, an atom index is used during force calculation in the final stage which we are not interested in.  Table~\ref{Tab1} summaries the loop indices.

\indTab

Keeping in mind the loop indices, we classified the parallelizable loops in a NAO DFT code.  The parts of effective potential calculation and diagonalization concern the Hamiltonian matrix and the overlap matrix, but the former part do not use k-point index.  These matrices are reducible and divided into respective k-point matrices (or spins).  So diagonalization can be easily parallelized within MPI because matrices of disparate k-points (or spins) require little information to communicate each other, i.e., close to embarrassing parallelism, which is one of reasons to make DFT codes MPI-friendly.  On the other hand, k-point index can be ignored in the effective potential calculation part because adding k-point dependence is just multiplying a simple matrix and consumes little time.  So MPI is not necessary in this part with respect to k-point index.  For basis function index, shared memory parallelization should be better to the Hamiltonian matrix and the overlap matrix than MPI because those matrices are irreducible after they are separated into single k-point ones (within non-spin or non-collinear spin calculation~\cite{19}, or with single spin within collinear spin calculation).  For auxiliary basis functions, consisting of density grids used in the parts of calculating effective potential or densities, shared memory parallelization looks more appropriate.

Since it is not easy to choose CUDA or OpenMP on the parts suitable to shared memory parallelism, we performed a simple test to compare CUDA and OpenMP.  Devices normalized --- divided every element of a vector by its norm --- n vectors which consist of n elements.  Parallelization decomposes this algorithm into n thread operations.  We utilized two Intel Xeon E5606 2.13GHz quad-core CPUs and one NVIDIA Quadro 4000 GPU.  OpenMP calculation used 4 or 8 CPU cores (i.e., 4 or 8 threads), while CUDA calculation used 1 GPU device with 256 cores.  Table~\ref{Tab2} shows how computation time changes as the number n increases.  CUDA calculation looks faster than others when n exceeds 1000.  On the other hand, the minima of OpenMP are always less than the maxima of CUDA because the speed of OpenMP or CUDA depends on the way of memory access.  Moreover, the parallelization with CUDA spends much more time whenever transferring between a GPU device and a CPU host.

\tetTab

Looking into the test result (Table~\ref{Tab2}) and Table~\ref{Tab1}, we extracted some guidelines for choice of shared memory parallelization methods.  It looks difficult even for faster CPUs to prevail over GPUs when n exceeds 1000, and it should be considered that Quadro 4000 is not that fast in comparison with other GPUs of compute capability 2.x~\cite{20}.  In consideration of large-scale calculations, since small-scale calculations do not need hybrid parallelization, CUDA should be preferred in most cases where coalesced memory access~\cite{21} is possible and the amount of data transferred between GPU and host is small compared to the calculation time of the data.  Otherwise, OpenMP should be chosen because its memory access requires locality~\cite{22,23,24,25,26,27}, as opposed to coalesced memory access.  However, if either the GPU (CUDA) or CPU (OpenMP or serial) parts continue too long, CPUs can take over a small portion of the CUDA-favored part or vice versa because GPUs can operate independently of CPUs.  

One can apply these rules to every part of DFT procedure, specifically.  Since each term in the effective potential can be calculated independently, one can apply CUDA and OpenMP alternately in the effective potential part, i.e., CUDA for the first term and OpenMP for the second term and so on.  On the other hand, the choice is not trivial for the density calculation part, but one can try dynamic load balancing between CUDA and OpenMP like Brown et al.~\cite{28}.  It is also difficult to settle the general choice for the diagonalization part because the choice should depend on its diagonalizing method, but we can obtain a hint from the fact that this part is full of matrix manipulation.  CUDA should be chosen if parallelizable loops of basis function index are two-fold or more in cases like matrix multiplication.  If such a parallelizable loop is just a single loop, the suitable shared memory parallelization method should be chosen to minimize data transfer between CPUs and GPUs, but it seems that CUDA should be preferred in most cases considering that DFT calculation scale will become larger and larger~\cite{note1}.
 
The above argument can be accompanied with MPI.  MPI can intervene in CUDA or OpenMP sections, especially in a calculation of effective potential or densities where MPI can be idle.  MPI can share loads with CUDA or OpenMP since the number of auxiliary basis functions is usually so large.  Then the calculation of effective potential part will be filled with rotations of MPI+CUDA and MPI+OpenMP, but to take advantage of simultaneous work of GPUs and CPUs, their sequence should be 

\begin{description}
\item[Step 1.] CUDA calculation
\item[Step 2.] OpenMP calculation
\item[Step 3.] MPI communication of OpenMP results
\item[Step 4.] Transferring CUDA results from GPUs to their hosts
\item[Step 5.] MPI communication of CUDA result
\end{description}

The step 4 can be done during the step 2 or 3 if asynchronous transfers~\cite{21} are possible.
The strategy for conversion from the MPI+CUDA 2-way hybrid parallelization to the 3-way hybrid parallelization is straightforward.  Looking into Tables~\ref{Tab1} and \ref{Tab2}, one can find that all OpenMP parts should convert into CUDA version unless data transfers between GPUs and CPUs are not severe.  If a data transfer is not negligible, the OpenMP part should return to its serial code form.  The strategy for effective potential part in the above paragraph can be modified by changing the step 2 with CUDA and removing the step 3.

\geoFig

\section{Application to O\lowercase{pen}MX}
We tested our strategies on OpenMX 3.6 code with a magnetite system.  OpenMX~\cite{2,29,30} was developed for quantum simulations of nano-scale materials and can deal with large-scale calculations on parallel computers, by implementing MPI and OpenMP.  Our magnetite system consists of 56 atoms (24 iron atoms and 32 oxygen atoms) in the calculation unit-cell (Fig. 2), and the atoms' coordinates and the unit-cell size were borrowed from X-ray measurement~\cite{32}.  Total numbers of k-points, main basis functions and auxiliary basis functions used in this system are 64, 1040 and 512000, respectively.  For this system, we performed spin-collinear band calculation within GGA.

\rouTab

Time consuming analysis of serial calculation of this system summarized in Table~\ref{Tab3} reveals that the diagonalization part possesses the most of the CPU time up to 94 \%.  Therefore, we focus only on the diagonalization which solves the eigenvalue equations.  In OpenMX, diagonalization within spin-collinear band calculation is governed by the routine called Band\_DFT\_Col, which consists of following parts.

\begin{description}
\item[Part 1.] Giving k-point dependence~\cite{note2} to overlap matrix S and Hamiltonian H
\item[Part 2.] Diagonalizing S to rearrange S
\item[Part 3.] Multiplication of matrices to make S$^\dagger$HS
\item[Part 4.] Diagnoalizing S$^\dagger$HS to get eigenvalues or eigenvectors
\item[Part 5.] Finding chemical potential and calculating band energy.
\item[Part 6.] Calculating charge density matrix and energy density matrix
\end{description}

Then the algorithm of Band\_DFT\_Col is, crudely expressed,
\\

\begin{tabular}{ll}
\textbf{for} & (assigned k-points) \\
{ } & Part 1 \\
{ } & Part 2 \\
{ } & Part 3 \\
{ } & Part 4 (obtaining eigenvalues) \\
\textbf{end} & \textbf{for} \\
Part & 5 \\
\textbf{for} & (assigned k-points) \\
{ } & Part 1 \\
{ } & Part 2 \\
{ } & Part 3 \\
{ } & Part 4 (obtaining eigenvectors) \\
\textbf{end} & \textbf{for} \\
Part & 6 \\
\end{tabular}
\\

The parts 2 and 4 use the Householder algorithm for tridiagonalization~\cite{34} and a LAPACK~\cite{35} routine among dstevx, dstedc or dstegr.  The parts 3 and 4 use ZGEMM routine in LAPACK~\cite{35}.  The parts 1, 5 and 6 are just multiplications and summations.  Table~\ref{Tab4} exhibits time sharing of each part for single CPU calculation.

\parTab

We could apply our strategy on each part separately.  The large loop with k-point index divided into MPI processes as mentioned in the previous section.  The ZGEMM routine was replaced by NVIDA's CUBLAS library~\cite{36} in CUDA toolkit 4.2.  In the part 1, CUDA manages overlap matrix S while the (OpenMP or serial) CPU code treats Hamiltonian H, but exchanging S and H is possible between the CPU code and the CUDA code.  The part 5 does not need parallelization because time sharing is very small.  CUDA has advantages also in the part 6 because the main multi loop of this part can be converted into a large loop of O(n2) where n is the number of basis functions. 

The parts 2 and 4 actually call the same subroutine named Eigen\_HH, composed of four steps: the Householder process, calling a LAPACK routine (dstevx, dstegr or dstedc), rearrangement of eigenvectors, and normalization.  The last two steps are activated only given an option demanding eigenvectors, and we chose CUDA for them because each of them has only one parallelizable loop but the amount of data transfers does not depend on whether CUDA or OpenMP is used.  Later, a test verified that CUDA is appropriate for them in case of calculating our magnetite system with 1040 basis functions~\cite{note1}.  The LAPACK routines, dstevx, dstegr and dstedc, could not be parallelized, but a small portion of the step rearranging eigenvectors can be calculated on GPUs while a LAPACK routine is working on CPUs. 

The remaining first step, tridiagonalization by Householder algorithm~\cite{34}, is an iterative loop that consists of following procedures~\cite{2,37}.  For the $i$th iteration ($i = 1, 2, \ldots , n-1$.  $n$: rank of the input matrix), where \textbf{B} is the matrix obtained from ($i-1$)th iteration (initially the input matrix),

\begin{description}
\item[Procedure 1.] Setting the vector \textbf{u} as ith column vector of \textbf{B}
\item[Procedure 2.] Calculating the scalar s (norm of the above vector \textbf{u}) and change the ith element of \textbf{u} by subtracting s.
\item[Procedure 3.] Storing the ith elements of a few vectors to be used in the rearrangement of eigenvectors
\item[Procedure 4.] Calculating the vector $\mathbf{p'} (= \mathbf{B} \cdot \mathbf{u} / (2\mathrm{u}^2)$ )  and the dot product of $\mathbf{p'}$ and \textbf{u}
\item[Procedure 5.] Calculating the vector $\mathbf{q'}$ from $\mathbf{p'}$ and $\mathbf{p' \cdot u}$ (See reference~\cite{37} for the definition of $\mathbf{q'}$)
\item[Procedure 6.] Partial transformation of \textbf{B} by subtracting a matrix composed of products of the \textbf{u} and $\mathbf{q'}$ elements. (See reference~\cite{37} for the form of the subtracting matrix)
\end{description}

The main loop of the Householder process in Eigen\_HH is repeating above procedures until the target matrix becomes a real symmetric tridiagonal matrix.  As our policy for matrix manipulation, we applied CUDA to procedure 6 which has a double parallelizable loop of basis function index.  The procedure 3, of no loop, naturally remained not parallelized.  Single parallelizable loop of such kind was found in each procedure 1, 2, 4, and 5, but we applied CUDA to 1, 4, and 5.  Due to the procedure 6, CUDA was the inevitable choice in the procedures 1 and 5 to avoid data transfer between CPUs and GPUs.  The procedure 2 and 4 includes summation, which leads to inefficiency in parallelization, but the procedure 4 can be performed in GPUs during the procedure 3.  (Dot Product calculation consists of multiplication and addition. We found that addition in dot product calculation is more efficient on CPUs even considering data transfer between CPUs and GPUs. So we assigned the addition in the procedure 4 to CPUs.)  So the only procedure 2 should be converted by OpenMP or should remain the serial form. 
After applying our strategies, we found that OpenMP fraction would be so small.  In our modified Band\_DFT\_Col, OpenMP was used in some initial settings, data type conversion of matrices, and the procedure 2 in Eigen\_HH.  All these occupy tiny portions of the whole Band\_DFT\_Col, which suggests that MPI+CUDA parallelization should reduce computational time much more than 3-way hybrid parallelization. Accidentally, these OpenMP parts are disadvantageous to be converted by CUDA. So MPI+CUDA 2-way parallelization is possible without CUDA conversion of these parts.

Even under our hybrid politics, comparing parallelizable fractions shows that OpenMX 3.6 DFT code prefers MPI to other parallelization methods.  MPI parallelized 99.5 \%, close to 100 \%, of the CPU single core time of Band\_DFT\_Col mainly by parallelization on k-points.  CUDA and OpenMP parallelized 88.2 \% of the CPU single core time.  Although it is difficult to distinguish OpenMP time and CUDA time completely due to GPUs' ability of independent working, we can approximate CUDA parallelizable fraction as 88 \% from our OpenMP time measurement close to 0.2 \%, but 88 \% is about 12 \% different from that of MPI.  According to Amdahl's law, if computation power is four times of CPU single core's, the result speedup by MPI will be about 3.9 while that by CUDA will be about 2.9.  In order to prevail over a quad-core CPU with MPI, a GPU with CUDA may need computational power more than about 6.5 times of CPU single core's.

Our machine setup is a single computer with two Intel Xeon E5606 processors and two NVIDIA Quadro 4000 graphics cards.  The main board of the machine has: two Intel Xeon E5606, each with 4 cores at 2.13Ghz, each and 12GB of DDR3 system memory that allows for a theoretical 17.06 GFLOPS performance in single core double precision floating point calculation~\cite{39}.  The memory test for Xeon E5606 single core was done by STREAM 5.10, and we obtained 6.48 GB/s as bandwidth.  Each of the two NVidia Quadro 4000 GPU processors has 256 cores and 2GB of GDDR5 memory with 89.6 GB/s bandwidth~\cite{40}.  The peak performance for one of our GPUs is 243.2 GFLOPS using double precision numbers~\cite{40}.  It is important to note that these are peak theoretical performances if all computations are performed with no latency or empty processing cycles.

Following the definition of speedup by Maintz et al.~\cite{8}, we estimated the ideal speed-up from above information if all operations were assumed to be done on a GPU.  If all operations were calculation, the speedup would be 14.26.  If all operations were memory access, it would be 13.83.  Both values are close to 14, but these values are not expectations of the actual speedup but just references to estimate the efficiency of CUDA parallelization because the ideal speedup calculation was based on theoretical peak performances and a theoretical bandwidth although CPU memory bandwidth is real one. 

\oneTab

We first used one quad-core CPU and one GPU, and the result is summarized in Table~\ref{Tab5}.  Since MPI parallelizes the code with respect to k-points, in case of MPI+CUDA parallelization, we allocated more k-points to GPUs, applying a weighting ratio of four, from the speedup by one GPU and one CPU core, denoted CUDA in Table~\ref{Tab5}.  Consequently, the best speedup was obtained by MPI+CUDA parallelization.  Although OpenMP's memory cost is about two times lower than MPI, it stands the worst in speedup, maybe due to memory bottleneck, low parallelizable fraction or cache miss.  CUDA enhances OpenMP's speedup well.  However, the speedups by CUDA do not reach our expectation.  Considering that Quadro 4000 is not that fast among compute capability 2.x~\cite{20} GPUs, larger speedups can be expected on other GPUs.

\parFig

We examined the speedups of the respective routines to find reasons of CUDA parallelization's unsatisfactory efficiency.  As shown by Fig. 3, parts 2 and 4 (diagnoalizing S and S$^\dagger$HS) are obstacles in increasing the whole speed.  The strangely high speedup of part 3 (making S$^\dagger$HS), mostly consisting of calling ZGEMM, implies that ZGEMM fits CUDA parallelization although the speedup over 14 seems to be caused by elimination of code lines unnecessary to CUDA parallelization.  ZGEMM shows the highest GPU speedup in Maintz et al.'s result as well~\cite{8}.  ZGEMM appears to be also the reason why part 4 (calling Eigen\_HH and ZGEMM each once) gains higher speedup than part 2.  Therefore, Eigen\_HH routine was suspected to be the main obstacle. 

\EHHTab

Thus we analyzed Eigen\_HH's performing time (Table~\ref{Tab6}).  Among the four steps of Eigen\_HH, the tridiagonalization and rearrangement of eigenvectors occupy about 85 \% of the single CPU core calculation time of Eigen\_HH, and their speedups by 1 GPU and 1 quad-core CPU were measured to be 5.23 and 4.23, respectively within CUDA+OpenMP parallelization.  Householder algorithm seems to be optimized to serial calculation, which appears to be the reason for the tridiagonalization's low speedup.  So further speed increases need a considerable change of algorithm, but that is beyond the scope of this paper.  The low efficiency of the rearranging eigenvectors is naturally expected since it has only single parallelizable loops.

\twoTab

Table~\ref{Tab7} shows our result for 2 quad-core CPUs and 2 GPUs.  Like the result for 1 CPU and 1 GPU, MPI+CUDA, our 2-way hybrid scheme yields the best speedup.  Our 3-way hybrid scheme stands the second, but its memory cost by CPUs was about 2.5 times lower than that of MPI+CUDA.  Speedup enhancement by CUDA is slightly more effective than the 1 CPU and 1 GPU cases.  Accuracy of our hybrid schemes is very similar to the non-CUDA cases since our hybrid schemes do not change the essence of the code's algorithm.

The good point in our hybrid schemes is that efficiency hardly drops as processors add.  By comparison of the 2 CPUs and 2 GPUs cases to the 1 CPU and 1 GPU cases (MPI to MPI, OpenMP to MPI+OpenMP, CUDA+OpenMP to 3-way hybrid, and MPI+CUDA to MPI+CUDA), the speed increase rates for the cases including CUDA are superior to the non-CUDA cases.  Each GPU is independent and interacts with the whole system only by its PCI-Express bus, which compensates the main memory bottleneck effect due to competition among CPU cores.  In MPI+CUDA scheme, one CPU core'sdifferent action may further compensate the bottleneck effect.

Our result reveals two features of hybrid parallelization with CUDA without changing the essence of the code's algorithm.  Large speedup enhancement in the cases with OpenMP suggests that GPUs' aid to parallelization is very useful to huge memory consuming calculation where few MPI processes per node are available. Little efficiency drop by processor addition implies the hybrid parallelization will be necessary to many multi-core CPU systems where adding CPUs will trigger large efficiency drop.  From both points, we can insist that even easy conversion with CUDA will be more or less helpful to DFT calculation on parallel computing systems.

\section{Conclusion}
We applied hybrid parallelization including CUDA to a MPI-friendly DFT source code, originally not optimized to CUDA.  Two types of hybridization are possible such as 3-way hybrid scheme saving memory with equally divided load and 2-way hybrid scheme maximizing speed by weighting of unequal loads.  We established several rules in editing NAO DFT codes to parallelize with hybrid schemes, by analyzing NAO DFT codes' property and comparing OpenMP and CUDA on a simple test.  Applying these rules is easy because they do not change the essence of the codes's algorithm.

The test was performed on a magnetite material system with OpenMX 3.6 code by utilizing a hardware system containing 2 Xeon E5606 CPUs and 2 Quadro 4000 GPUs.  3-way hybrid scheme obtained the speedup of 7.55 and saved nearly 60 \% of memory.  2-way hybrid scheme obtained the speedup of 10.94.  CUDA did not reach our expectation, due to the relatively low CUDA parallelizable fraction of the code and its algorithm's inefficiency for CUDA, compared with MPI.  Nevertheless, CUDA capable GPUs complement the efficiency of OpenMP and compensate the CPUs' excessive competition for memory in MPI.

GPGPU appears meaningful even by simple conversion within hybrid parallelization with a not-so-fast GPU.  Note that we assumed compute capability 2.x~\cite{20}, and a change of GPU can give better performance because the Quadro 4000 is only a midrange GPU when compared with the NVIDA Tesla¡¯s.  If one wants to further speed up the code and has a strong will to meticulously edit the code, one can change the essential algorithms of a DFT code not immediately CUDA-friendly.

\section*{ACKNOWLEDGEMENT}
We thank Prof. T. Ozaki for allowing us to edit OpenMX 3.6 code.  This work was supported by the Technology Innovation Program (10036459, Development of center to support QoLT industry and infrastructures) funded by the MOTIE/KEIT and the National Research Foundation of Korea (NRF) grant funded by the Korea government (MEST. No. 2009-0092790).


\begin{thebibliography}{99}
\bibitem {note1} One can program the code to choose whether CUDA or OpenMP, depending on matrix sizes, but its criterion on machines.
\bibitem{note2} This part belongs to the part of effective potential calculation, but Band\_DFT\_Col includes it in OpenMX.
\end{thebibliography}
\end{document}